\documentstyle[12pt,psfig]{article}

\def\Z{{\cal Z}}
\def\t{\tilde}
\def\h{\hat}
\def\th-vacuum{$\Theta$-vacuum}
\def\be{\begin{equation}}
\def\ee{\end{equation}}

\newcommand{\AmS}{{\protect\the\textfont2
\renewcommand{\thesection}{\Roman{section}}
  A\kern-.1667em\lower.5ex\hbox{M}\kern-.125emS}}
 
\begin{document}
\rightline {DFTUZ 99/10}
\vskip 2. truecm
\centerline{\bf Vector Symmetries and $\Theta$-Vacuum}
\vskip 2 truecm
\centerline { Vicente ~Azcoiti and Angelo ~Galante}
\vskip 1 truecm
\centerline {\it Departamento de F\'\i sica Te\'orica, Facultad 
de Ciencias, Universidad de Zaragoza,}
\centerline {\it 50009 Zaragoza (Spain).}
\vskip 3 truecm

\centerline {ABSTRACT}
\vskip 0.5truecm

\noindent
Under very general assumptions we show that Vafa-Witten theorem
on vector symmetries in vector-like theories can be extended to
some physically relevant gauge theories with non-positive
definite integration measure as $QCD$ with a \th-vacuum term.

\vfill\eject
\baselineskip=24pt

The vacuum realization of vector and axial symmetries in gauge
theories with dynamical fermions, as $QCD$, plays a fundamental
role in the understanding of the dynamics of these models at
zero and finite temperature. The spontaneous breakdown of 
chiral symmetry in $QCD$ at zero temperature and the realization
of the Goldstone theorem allow us to understand many essential
features of the low energy spectrum of this model, and the
restoration of this symmetry at high temperature is the main
signal expected for the deconfining phase transition.

Concerning vector symmetries as flavor or 
baryon number, Vafa and Witten showed \cite{witten} few years
ago that they can not be spontaneously broken in
vector-like theories as $QCD$. The two main ingredients in
the demonstration were $i)$ the positivity of the path integral measure
in the effective gauge theory obtained after integration of the
Grassman degrees of freedom and $ii)$ the antihermiticity and local 
character of the massless Dirac operator.

The second ingredient was essential in order to derive bounds 
which show exponential fall-off at large distances
for the massive quark propagator in any background gauge field.
The authors of \cite{witten} noticed that
the correlation function of any quark-composite local 
gauge-invariant operator carrying a non vanishing conserved vector
charge should be proportional to at least one power of the
quark propagator. Taking also into account that the mean value
of an upper-bounded quantity computed with a probability
distribution function is always less or equal than its upper-bound,
an exponential fall-off at large distances  was found for 
the correlation function of all vector charge carrying operators. 
This result allowed to show not only that the mean value of any
local order parameter for a vector symmetry should vanish but
also that massless quark composites can not be constructed from
massive constituents.

Even if Vafa-Witten theorem applies also to vector-like theories
at finite temperature, there are some physically relevant
formulations in which at least one of the two 
essential ingredients in the demonstration are not realized.
The most interesting cases are $QCD$ with a \th-vacuum term 
and $QCD$ at finite chemical potential.

In $QCD$ with a \th-vacuum term the factor of $i$ picked-up by the
$F\t{F}$ operator under Wick rotation makes the 
euclidean action complex. Even if the partition function is
real, the integration measure is no longer positive definite
and therefore the first ingredient in the theorem breaks-down.
The second one is however preserved since the \th-vacuum term
does not change the structure of the massless Dirac operator.

The introduction of a finite chemical potential is $SU(N)$
gauge theories  makes also the effective gauge action obtained
after integration of the Grassman degrees of freedom complex
in the most interesting case of $N=3$.
Furthermore the introduction of a chemical potential modifies
also the structure of the massless Dirac operator. It looses
its antihermiticity and the second ingredient of the theorem
breaks-down too.

Since both cases, $QCD$ with a \th-vacuum term and
$QCD$ at finite chemical potential, are physically relevant,
it is worthwhile to ask to which extent and under what limitations
the theorem can be extended to these formulations.
We will devote this paper to $QCD$ with a \th-vacuum term 
leaving the finite chemical potential case,
which shows the additional complication previously stated,
for a separate publication.

In order to extend the theorem to $QCD$ with
a \th-vacuum term we will made use of two very general
assumptions. First we will assume that Parity is not
spontaneously broken in $QCD$ at $\theta=0$, which is a
necessary condition for the euclidean free energy density
be well defined at $\theta\neq 0$ \cite{paridad}.

The second general assumption, which is standard in statistical
mechanics, is that in the thermodynamical limit any intensive
operator does not fluctuate in a vacuum or equilibrium state. 
This is equivalent to the general  Quantum Field Theory
statement that all connected correlation functions of local operators 
verify cluster.

To better understand the physical meaning of this last assumption
let us think for a while in the Ising model.
In the high temperature symmetric phase the system has one
vacuum or equilibrium state. The density of magnetization is an
intensive operator and its probability distribution function will
become a single $\delta$ function in the thermodynamical 
limit since otherwise we would have fluctuations of the
density of magnetization. Since the $Z(2)$ symmetry of the Ising 
model is realized in the high temperature phase, the $\delta$ 
function is centered at the origin.

In the low temperature phase the $Z(2)$ symmetry is
spontaneously broken. Since the density of magnetization
$m_I$ is an order parameter for this symmetry, its
probability distribution function becomes now $1/2$ 
the sum of two $\delta$ functions. 
The density of magnetization fluctuates in the
thermodynamical limit because we have two vacuum or
equilibrium states and $m_I$ is not invariant under $Z(2)$
transformations. However if we choose an intensive operator 
invariant under $Z(2)$ transformations, as for instance the
energy density or $m_I^2$, it does not fluctuate in spite of the fact
that we have two vacuum states. Indeed the mean value of any
power of these operators does not depend on the vacuum state
it is computed.

The lesson we learn from this simple example is that if
the only reason to have a degenerate vacuum is spontaneous 
symmetry breaking ($i.e.$ no accidental vacuum degeneration)
and we choose an intensive operator 
invariant under symmetry transformations, it does not
fluctuate in the thermodynamical limit independently of 
the symmetry realization.
These features are not exclusive of the Ising model
but a general property of any statistical system.

Let us consider now the euclidean formulation of $QCD$ with
a \th-vacuum term. The euclidean partition function is
\be\label{2}
\Z = \int [dA_\mu][d\bar\psi][d\psi]
e^{ -\int d^4x \left({\cal L}(x)-\frac{i\theta}{16\pi^2} X(x)\right)}
\ee
where ${\cal L}(x)$ is the standard $QCD$ lagrangian and
$X(x)=\epsilon_{\mu\nu\alpha\beta}Tr F_{\mu\nu}F_{\alpha\beta}$
the \th-vacuum lagrangian. $X(x)$ is real in euclidean space
and we have exhibited in (\ref{2}) the factor of $i$ which
arises from Wick rotation.

When we integrate out the fermionic degrees of freedom we get an 
effective gauge theory
\be\label{3}
\Z = \int [dA_\mu] \det\Delta(m,A_\mu)e^{-S_G(A_\mu)}e^{i\theta\h X(A_\mu)} ,
\ee
where $\Delta(m,A_\mu)=m+i\Lambda(A_\mu)$ is the Dirac operator for quarks
of bare mass $m$, $S_G(A_\mu)$ the pure gauge action and
$\h X(A_\mu)=\frac{1}{16\pi^2}\int d^4x X(x)$ the \th-vacuum action.

The integrand in (\ref{3}) is a complex number due to the \th-vacuum
contribution. 
The imaginary part of the integrand gives no contribution to the 
partition function (the $QCD$ action is parity invariant). 
The real part however is not positive definite except at $\theta=0$.

Let $J(x)$ be any quark composite gauge invariant operator 
with non zero isospin or baryon number and let us consider the
correlation function $Q(x)=\langle J(x)J^*(0)\rangle$.
This correlation function will be given by the following ratio
\be\label{4}
Q(x)=
\frac{
\int [dA_\mu] Q_A(x)\det\Delta(m,A_\mu)e^{-S_G(A_\mu)+i\theta\h X(A_\mu)}
}
{
\int [dA_\mu] \det\Delta(m,A_\mu)e^{-S_G(A_\mu)+i\theta\h X(A_\mu)}
}
\ee
where $Q_A(x)$ is the correlation function in a given background
gauge field $A_\mu(x)$. If, as previously assumed, $J(x)$ carries a
conserved fermion charge, then $Q_A(x)$ will be proportional 
to at least one power of the fermion propagator 
$\Delta^{-1}_A(m,0,x)$ from 0 to $x$.
Vafa and Witten essentially showed in \cite{witten} that the
norm of the quark propagator matrix
\be\label{5}
\|\Delta^{-1}_A(m,0,x)\| \leq \alpha e^{-\beta\|x-y\|}
\ee
falls-off at least exponentially for any gauge configuration
since $\alpha$ and $\beta$ in (\ref{5}) are two constants
independent of the gauge configuration. In the previous statement
the world essentially stands for the fact that, strictly speaking,
relation (\ref{5}) is not true in the continuum formulation
(it is true in the lattice regularization approach).
This problem can be surmounted by considering smeared correlation
functions \cite{witten} or using a lattice regularization scheme.
In order to preserve as much as possible the 
simplicity in the notations, we will assume relation (\ref{5})
true in what follows.

At $\theta=0$, relation (\ref{5}) plus the fact that the
integration measure in (\ref{4}) defines a probability
distribution function, carried Vafa and Witten to conclude 
that a bound similar to (\ref{5}) holds also for the
correlation function $Q(x)$ and consequently that vector
symmetries can not be spontaneously broken. They noticed also
in \cite{witten} that introduction of a \th-vacuum term
would invalidate the proof because the positivity of the
integration measure is loosed.

To overcome this problem, we will make use now of the second
general assumption discussed in the introduction of this paper,
$i.e.$ in any statistical system intensive operators do not fluctuate
or equivalently, in a well defined quantum system, all the correlation
functions of local operators verify the cluster property.
We will apply this general assumption to the intensive operator 
$\t X(A_\mu) = \frac{1}{V}\frac{1}{16\pi^2}\int d^4x \h X(A_\mu)$
which defines the \th-vacuum term in the action.

The cluster property tells us that
\be\label{6}
\left\langle \t{X}^n \right\rangle = \left\langle \t{X} \right\rangle^n
\qquad\qquad\qquad \forall n
\ee

Due to the factor of $i$ which arises from Wick rotation, the
mean value $<\t{X}>$ is a pure imaginary number whereas the operator
$\t{X}$ takes real values in euclidean space.
This two features however are not in contradiction since the mean 
value in (\ref{6}) is computed with a complex action. In order to
avoid complications related to factors of $i$, we will consider the
operator 
\be\label{7}
Y = \t{X}^4 ,
\ee
the mean value of which is always real and positive.
Furthermore (\ref{6}) implies analogous relations for the $Y$ 
operator:
\be\label{8}
\left\langle Y^n \right\rangle = \left\langle Y \right\rangle^n
\qquad\qquad\qquad \forall n
\ee

Let us assume now that $O$ is any parity conserving operator
as for instance the operator $Y$ previously defined or the 
correlation function of any  gauge invariant quark-composite
operator driving flavor or baryon number.
We can write
\be\label{9}
\langle O\rangle=
\frac{
\int [dA_\mu][d\bar\psi][d\psi] O(\bar\psi,\psi,A_\mu)
e^{-S_{QCD}(\bar\psi,\psi,A_\mu)}\cos(\theta V\t{X})
}
{
\int [dA_\mu][d\bar\psi][d\psi] 
e^{-S_{QCD}(\bar\psi,\psi,A_\mu)}\cos(\theta V\t{X})
}
\ee
which, after integration of the Grassman degrees of freedom
becomes
\be\label{10}
\langle O\rangle =
\frac{
\int [dA_\mu] O(A_\mu)\det\Delta(A_\mu)e^{-S_G(A_\mu)}\cos(\theta V\t{X})
}
{
\int [dA_\mu] \det\Delta(A_\mu)e^{-S_G(A_\mu)}\cos(\theta V\t{X})
}
\ee
where, in order to leave notation as simple as possible, we have
used the same symbol $O$ to design the operator in the effective
gauge theory obtained after integration of the fermion degrees
of freedom.

In order to define a pseudo-probability distribution function we
will introduce in (\ref{10}) the trivial identity
\be\label{11}
1 = \int_0^\infty d\bar Y \delta(Y(A_\mu)-\bar Y)
\ee
which allows us to write expression (\ref{10}) as
\be\label{12}
\langle O\rangle=
\frac{
\int_0^\infty d\bar Y N(\bar Y) O(\bar Y) e^{-S_{eff}(\bar Y)}
\cos(\theta V\bar{Y}^{1/4})
}
{
\int_0^\infty d\bar Y N(\bar Y) e^{-S_{eff}(\bar Y)}
\cos(\theta V\bar{Y}^{1/4})
}
\ee
where $N(\bar Y)$ is the density of states
\be\label{13}
N(\bar Y) = \int [dA_\mu] \delta(Y(A_\mu)-\bar Y)
\ee
and $e^{-S_{eff}(\bar Y)}$ and $O(\bar Y)$ are respectively
$$
e^{-S_{eff}(\bar Y)} = 
\frac{
\int [dA_\mu] \det\Delta(A_\mu)e^{-S_G(A_\mu)}\delta(Y(A_\mu)-\bar Y)
}
{
\int [dA_\mu] \delta(Y(A_\mu)-\bar Y)
} 
$$
\be\label{14}
O(\bar Y) = 
\frac{
\int [dA_\mu] \det\Delta(A_\mu)e^{-S_G(A_\mu)}O(A_\mu)
\delta(Y(A_\mu)-\bar Y)
}
{
\int [dA_\mu] \det\Delta(A_\mu)e^{-S_G(A_\mu)}\delta(Y(A_\mu)-\bar Y)
}
\ee

This relations allow us to write for $<O>$ the simple expression
\be\label{15}
\langle O\rangle=
\int_0^\infty d\bar Y P(\bar Y) O(\bar Y) 
\ee
where 
\be\label{16}
P(\bar Y) =
\frac{
 N(\bar Y) e^{S_{eff}(\bar Y)} \cos(\theta V\bar{Y}^{1/4})
}
{
\int_0^\infty d\bar Y N(\bar Y) e^{S_{eff}(\bar Y)}
\cos(\theta V\bar{Y}^{1/4})
}
\ee
is a pseudo-probability distribution function since it is not
positive definite due to the factor $\cos(\theta V\bar{Y}^{1/4})$ in
the numerator. In spite of that, equations (\ref{8}) and (\ref{15})
tell us that in the infinite volume limit 
\be\label{17}
\int_0^\infty d\bar Y P(\bar Y) \bar{Y}^n = \bar{Y}_0^n
\qquad\qquad  \bar{Y}_0^n \geq 0
\ee

Equation (\ref{17}) implies that in the infinite volume limit
$P(\bar Y)$ becomes a true probability distribution function:
a Dirac distribution $\delta(\bar Y - \bar{Y}_0)$.
The mean value of operator $O$ will be in this limit
\be\label{18}
\langle O\rangle=
\int_0^\infty d\bar Y \delta(\bar Y - \bar{Y}_0)  = O(\bar{Y}_0)
\ee
where $\bar Y_0$ will depend on the theory parameters: gauge coupling,
quark masses and $\theta$ angle.

It turns out that to compute the expectation value of our
parity invariant operator it is enough to calculate its
average over a subset of gauge fields, actually
all the gauge fields characterized by a fixed
value of $Y(A_\mu)$. The expectation value of our operator coincides with its
mean value given by the expression (\ref{14}) and evaluated at 
$\bar Y =\bar{Y}_0$.

Let's see the consequences for any correlation function of a gauge 
invariant composite quark operator driving flavor or baryon number.
It is clear that expression (\ref{14}) defines a true
probability distribution function (the integration measure is 
positive definite for every $\bar Y$).
It follows that if $|O(A_\mu)|$ verifies the Vafa-Witten bounds
for all gauge configurations the same happens for $|O(\bar Y)|$
whatever the value of $\bar Y$. 
In particular this will be true for $\bar Y=\bar Y_0$ and therefore
we conclude that the expectation value of any of these correlation 
functions in presence of a non zero \th-vacuum term shows exponential 
fall-off at large distances.

In conclusion the two main consequences of Vafa-Witten theorem
$i)$ the impossibility to break spontaneously a vector symmetry
and $ii)$ the non existence of massless bound states made up with
massive constituents, apply also to $QCD$ with a \th-vacuum
term, assumed the euclidean formulation of this model can be
consistently done.

\noindent
{\bf Acknowledgements}
\vskip 0.3truecm

This work has been partially supported by CICYT (Proyecto AEN97-1680). 
A.G. was supported by a Istituto Nazionale di Fisica Nucleare fellowship
at the University of Zaragoza.


\newpage
\vskip 1 truecm

\begin{thebibliography}{9}

\bibitem{witten}
C. Vafa, E. Witten, 
Nucl. Phys. {\bf B234} \rm (1984) 173. 

\bibitem{paridad}
V. Azcoiti, A. Galante, hep-th/9901068.  

\end{thebibliography}
\end{document}